\documentclass[preprint]{iucr} 

     \journalcode{S}              
\usepackage{url}
\usepackage{graphicx}
\usepackage{float}
\usepackage{tabularx}

\begin{document}                  



\title{The Time-resolved Atomic, Molecular and Optical Science Instrument at the Linac Coherent Light Source}
\shorttitle{TMO}


\cauthor[a]{Peter}{Walter}{pwalter@slac.stanford.edu}{address if different from \aff}
\author[a]{Timur}{Osipov}
\author[a]{Ming-Fu}{Lin}
\author[a, b]{James}{Cryan}
\author[a]{Taran}{Driver}
\author[a]{Andrei}{Kamalov}
\author[a]{Agostino}{Marinelli}
\author[a]{Joe}{Robinson}
\author[a]{Matt}{Seaberg}
\author[a, b]{Thomas J. A.}{Wolf}
\author[a]{Jeff}{Aldrich}
\author[a]{Nolan}{Brown}
\author[b]{Elio G.}{Champenois}
\author[a]{Xinxin}{Cheng}
\author[a]{Daniele}{Cocco}
\author[a]{Alan}{Conder}
\author[a]{Ivan}{Curiel}
\author[a]{Adam}{Egger}
\author[a]{James M.}{Glownia}
\author[a]{Philip}{Heimann}
\author[a]{Michael}{Holmes}
\author[a]{Tyler}{Johnson}
\author[a]{Xiang}{Li}
\author[a]{Stefan}{Moeller}
\author[a]{Daniel S}{Morton}
\author[a]{May Ling}{Ng}
\author[a]{Kayla}{Ninh}
\author[a,b]{Jordan T.}{O’Neal}
\author[a]{Razib}{Obaid}
\author[a]{Allen}{Pai}
\author[a]{William}{Schlotter}
\author[a]{Jackson}{Shepard}
\author[a]{Niranjan}{Shivaram}
\author[a]{Peter}{Stefan}
\author[a]{Xiong}{Van}
\author[a,b]{Anna Li}{Wang}
\author[a]{Hengzi}{Wang}
\author[a]{Jing}{Yin}
\author[a]{Sameen}{Yunus}
\author[a]{David}{Fritz}
\author[a]{Justin}{James}
\author[a]{Jean-Charles}{Castagna}

\aff[a]{SLAC National Accelerator Laboratory, 2575 Sand Hill Road, Menlo Park, California 94025, \country{USA}}
\aff[b]{Stanford PULSE Institute, SLAC National Accelerator Laboratory, 2575 Sand Hill Road, Menlo Park, CA 94025, \country{USA}}









\maketitle                        

\begin{abstract}
The newly constructed Time-resolved atomic, Molecular and Optical science instrument (TMO), is configured to take full advantage of both linear accelerators at SLAC National Accelerator Laboratory, the copper accelerator operating at a repetition rate of 120 Hz providing high per pulse energy, as well as the superconducting accelerator operating at a repetition rate of about 1 MHz providing high average intensity. Both accelerators build a soft X-ray free electron laser with the new variable gab undulator section. With this flexible light sources, TMO supports many experimental techniques not previously available at LCLS and will have two X-ray beam focus spots in line. Thereby, TMO supports Atomic, Molecular and Optical (AMO), strong-field and nonlinear science and will host a designated new dynamic reaction microscope with a sub-micron X-ray focus spot. The flexible instrument design is optimized for studying ultrafast electronic and molecular phenomena and can take full advantage of the sub-femtosecond soft X-ray pulse generation program.

\end{abstract}

\section{Introduction}

\begin{figure}
\centering
\includegraphics[width=1\textwidth]{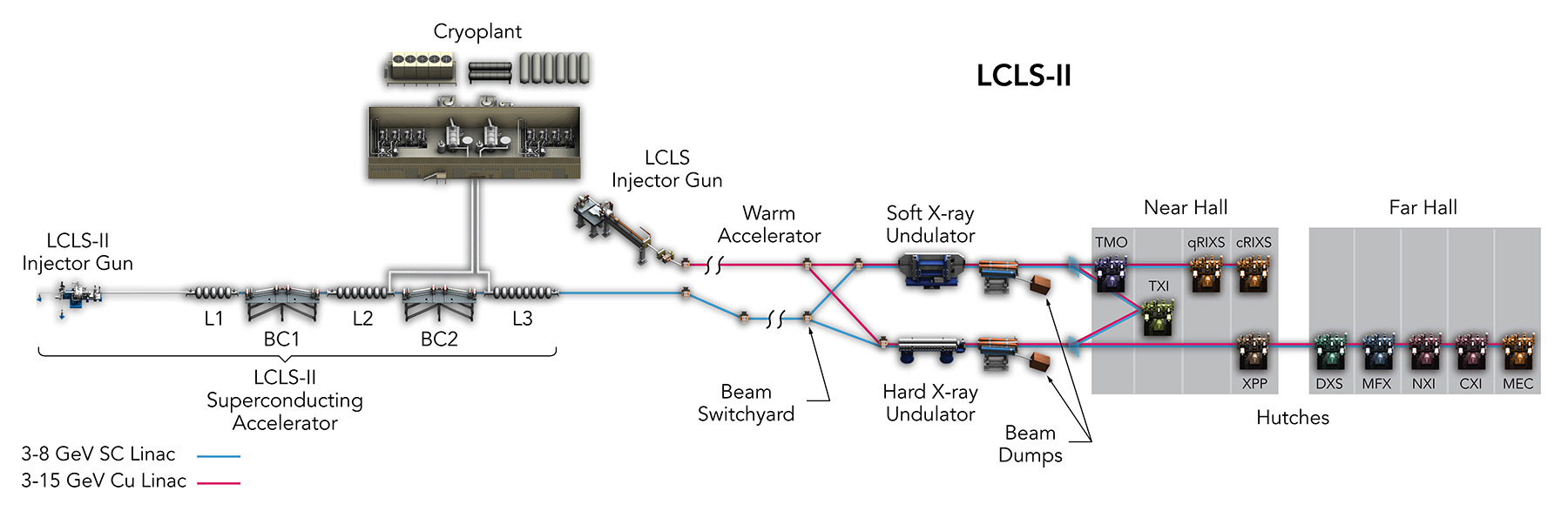}
\label{fig:LCLS}
\caption{Schematic overview of the new LCLS layout with both accelerators (LCLS(red) and LCLS-II(blue)). The LCLS-II X-ray laser is shown alongside the existing LCLS. LCLS uses the last third of SLAC's 2-mile-long linear accelerator - a hollow copper structure that operates at room temperature and allows the generation of 120 X-ray pulses per second. For LCLS-II, the first third of the copper accelerator has been replaced with a superconducting one, capable of a repetition rate of up to 1 MHz. Future Far Hall layout changes for LCLS II HE are not included.}
\end{figure}

The unique capabilities of LCLS, \cite{Arthur2002, Emma2010} the world’s first hard X-ray Free Electron Laser (FEL), have had significant impact on advancing our understanding across a broad range of scientific fields, from fundamental atomic and molecular physics, to condensed matter, catalysis, and structural biology. \cite{Young2010, Seibert2011, Chapman2011, Gomez906, Bostedt2016, Hartmann2018, ONEAL, Ilchen2021} A major upgrade of the LCLS facility, the LCLS-II project, is now underway. LCLS-II is being developed as a high repetition rate FEL. It features a 4 GeV continuous wave superconducting linac that is capable of producing uniformly spaced (or programmable) ultrafast X-ray laser pulses at a repetition rate up to 1 MHz spanning the energy range from 0.25 to 5 keV. Fig \ref{fig:LCLS} shows a schematic overview of the new FEL layout with both accelerators LCLS and LCLS-II. \\
To make the performance of the new high repetition rate FEL available for three new instruments, the existing Near Experimental Hall (NEH) of LCLS was completely reconfigured and hosts two new soft X-ray instrument and one tender X-ray instrument. The first of these instrument, the Time-resolved atomic, Molecular, and Optical Science (TMO) instrument is available now for user experiments. In this publication we will provide an overview of the TMO instrument including the various endstations, X-ray optics, optical laser systems and available detectors. We conclude with a brief description of the main scientific research possible in TMO and some first results.  
\section{Instrument overview}

\begin{figure}
\caption{Schematic overview of the TMO instrument layout with distances. All distances are indicated in meters from the first mirror inside of TMO. Shown are scatter Slits (S), Diagnostics (D), KB Optics (KBO), Laser IN-coupling (L-IN), Interaction Points (IP), Laser OUT-coupling (L-OUT) and laser Arrival Time Monitor (ATM).}
\includegraphics[width=1\textwidth]{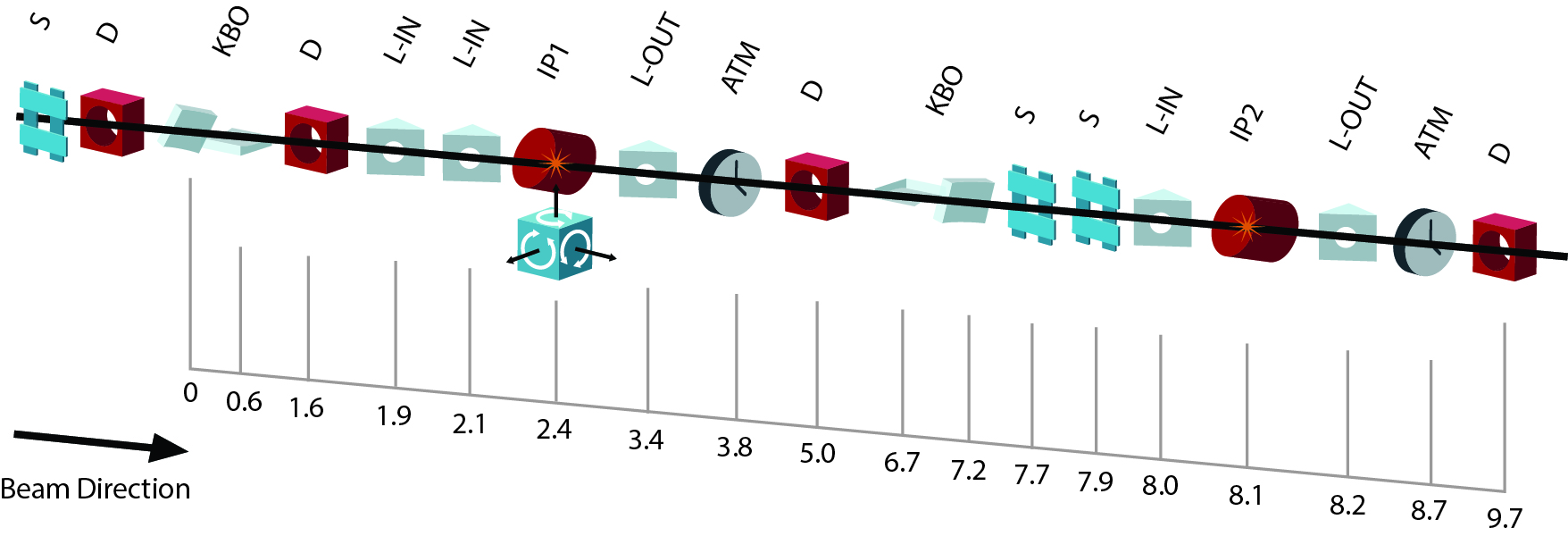}
\label{fig:TMO2}
\end{figure}

\begin{figure}
\caption{Overview of the TMO instrument layout. Showing both endstations which are capable to take full advantage of both the high per pulse energy from the copper accelerator (120 Hz) as well as high average intensity and high repetition rate from the superconducting accelerator (1MHz). Indicated are the beam position monitors IM3K4, IM4K4, IM5K4, IM6K4; differential pumping sections PA1K4 and PA2K4; IP1 optical laser in-coupling LI1K4, LI2K4 and IP1 laser out-coupling LI3K4; the arrival time monitors TM1K4 and TM2K4; wave front sensor PF1K4 and PF2K4; retractable beam terminators SF1K4 and SF2K4; beamline collimators PC5K4; as well as the interaction points IP1 and IP2; and finally the beam terminator SF3K4.}
\includegraphics[width=1\textwidth]{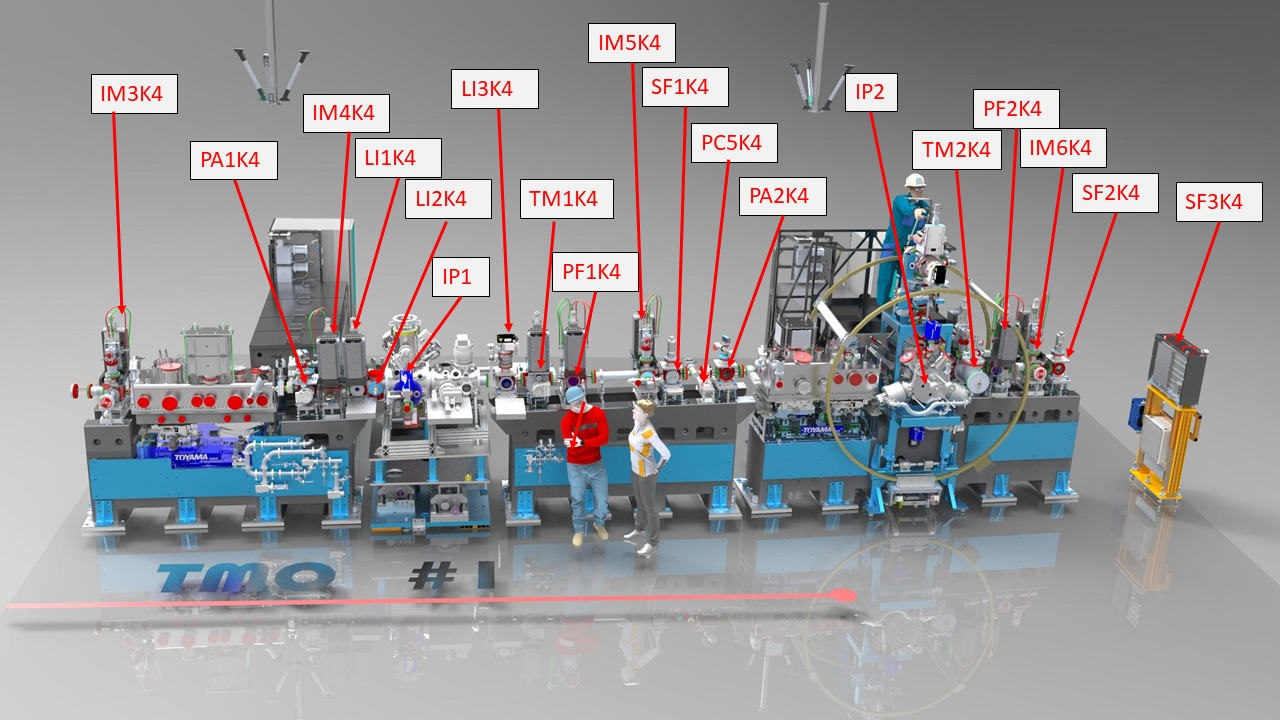}
\label{fig:TMO}
\end{figure}

To guarantee a high flux soft-X-ray beam line, TMO stays in the NEH basement hutch formerly occupied by the Atomic and Molecular Optics (AMO) instrument (see Fig. \ref{fig:LCLS}), [\cite{Bozek2009, Ferguson2015}] and has only one horizontal mirror between the undulators and the hutch (see Fig \ref{fig:KBO}). All devices such as diagnostics, apertures and optics are designed such that TMO can be operated in the energy range from 0.25 - 2.2 keV and 0.25 - 1.4 keV, respectively, and can take full advantage of both accelerators and the variable gap soft X-ray undulators. \cite{Wallen2016} The two different energy ranges result from the two different reflection cut offs of the two different mirror systems. The high peak power copper accelerator and the high repetition rate superconducting accelerator operating at 120 Hz and up to 1 MHz, respectively. The NEH 1.1. (TMO) instrument has two separate focus spots (i.e. interaction points) in series (see Fig.\ref{fig:TMO2}, \ref{fig:TMO}, \ref{fig:KBO}). Each focus spot has a separate Kirkpatrick-Baez (KB)-mirror \cite{Baez1946} system, focus diagnostic, sample delivery environment, laser module table and controls system. Consequently, each focus spot can be operated independently but also simultaneously. The following subsection will give a more detailed description of TMO. An overview of the TMO instrument layout is given in Fig.\ref{fig:TMO2}, \ref{fig:TMO}.
\subsection{X-ray optics}
\begin{figure}
\caption{Overview of the TMO mirror layout. (a) shows the distances of each mirror and IP from the undulator in meter. The shown section inside the dashed line marks the TMO hutch with the KBO systems. (b) shows the dashed line section of the top part with the mirror orientations and deflection angle as top and side view. (c) shows the transmission (T) over the photon energy (E) for IP1 with the calculated average transmission of 80$\%$ (orange dash line) for both mirror coatings and (d)  shows the transmission (T) over the photon energy (E) for IP2 with the calculated average transmission of 60$\%$ (orange dash line) for both mirror coatings.}
\includegraphics[width=1\textwidth]{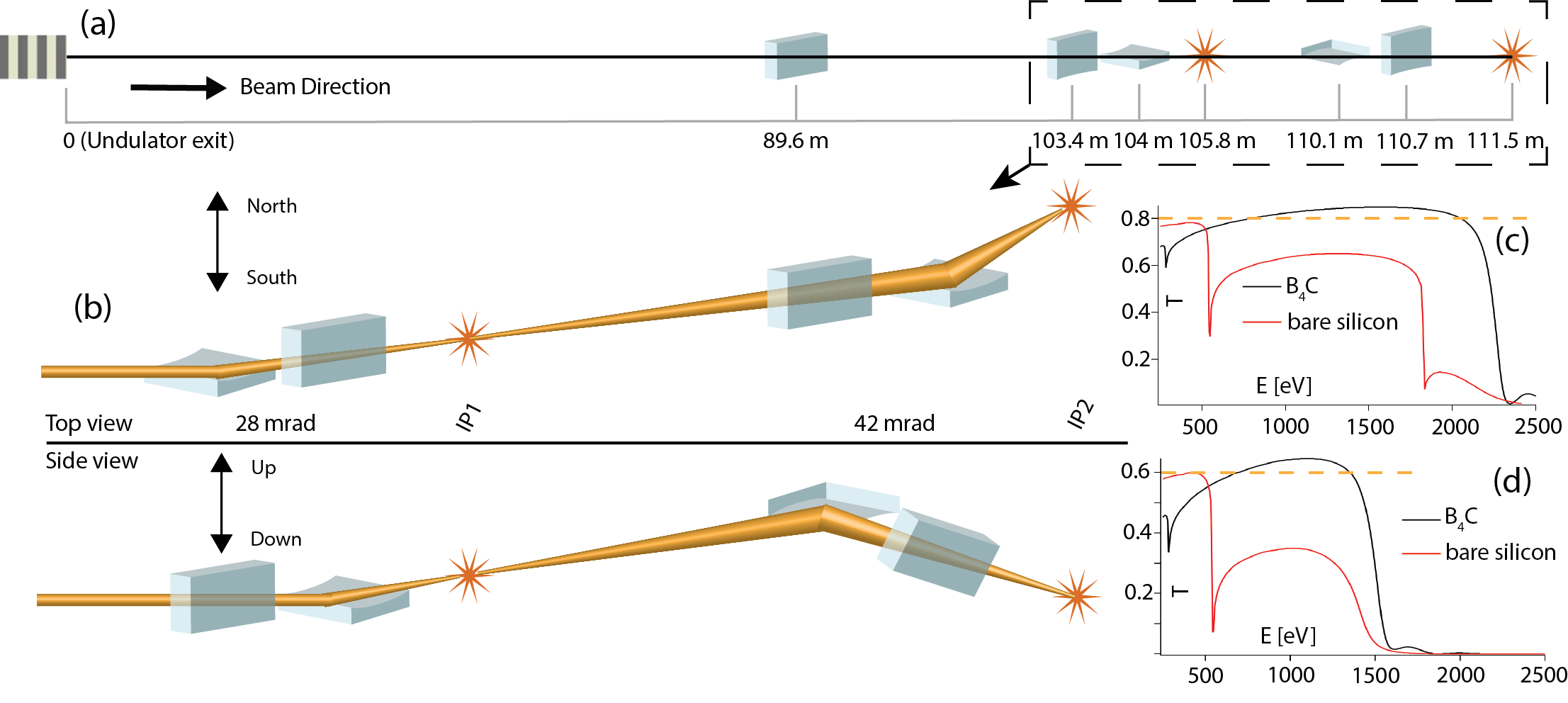}
\label{fig:KBO}
\end{figure}

\begin{figure}
  \centering
  \begin{minipage}[b]{0.49\textwidth}
    \includegraphics[width=\textwidth]{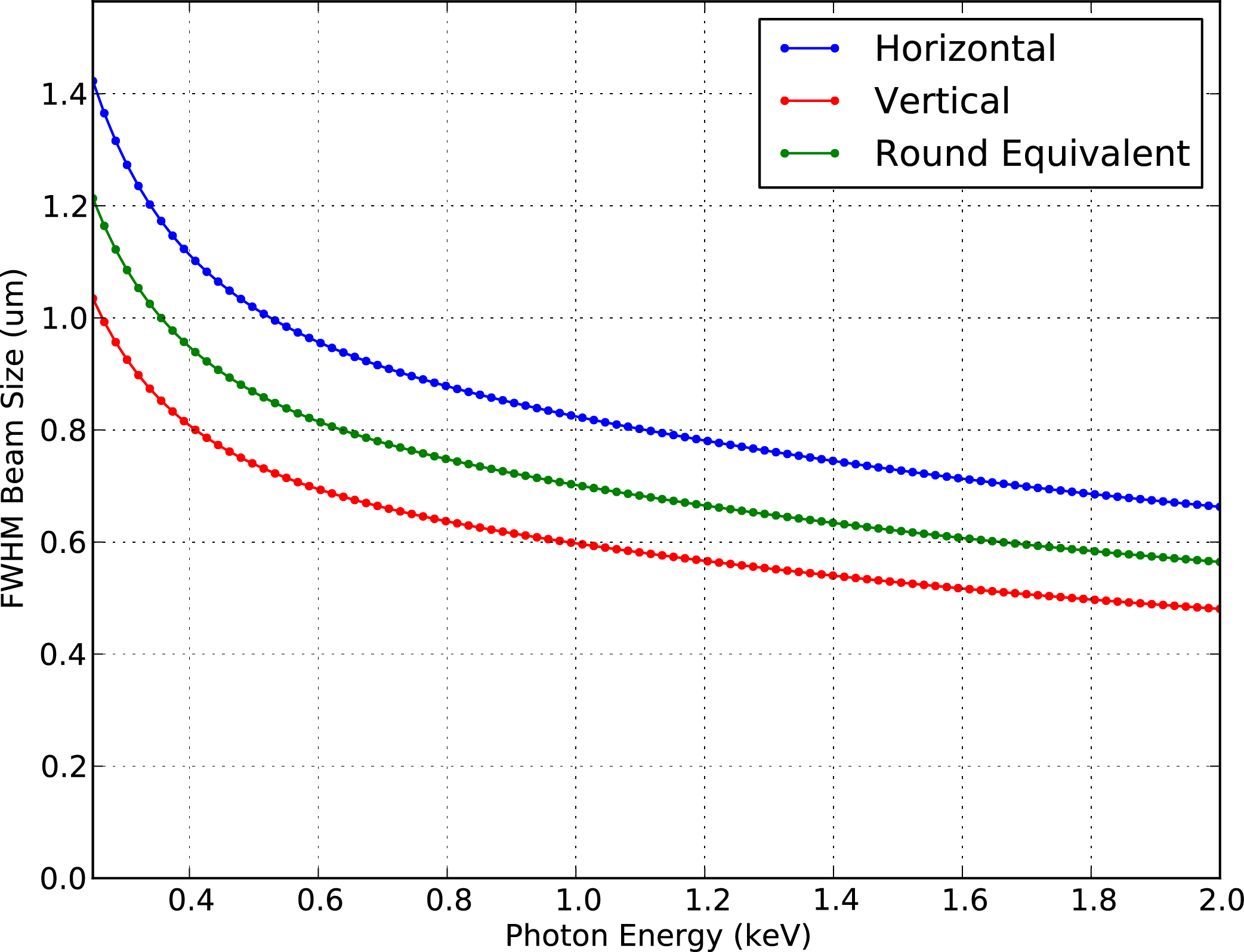}
    \caption{Flower one.}
  \end{minipage}
  \hfill
  \begin{minipage}[b]{0.49\textwidth}
    \includegraphics[width=\textwidth]{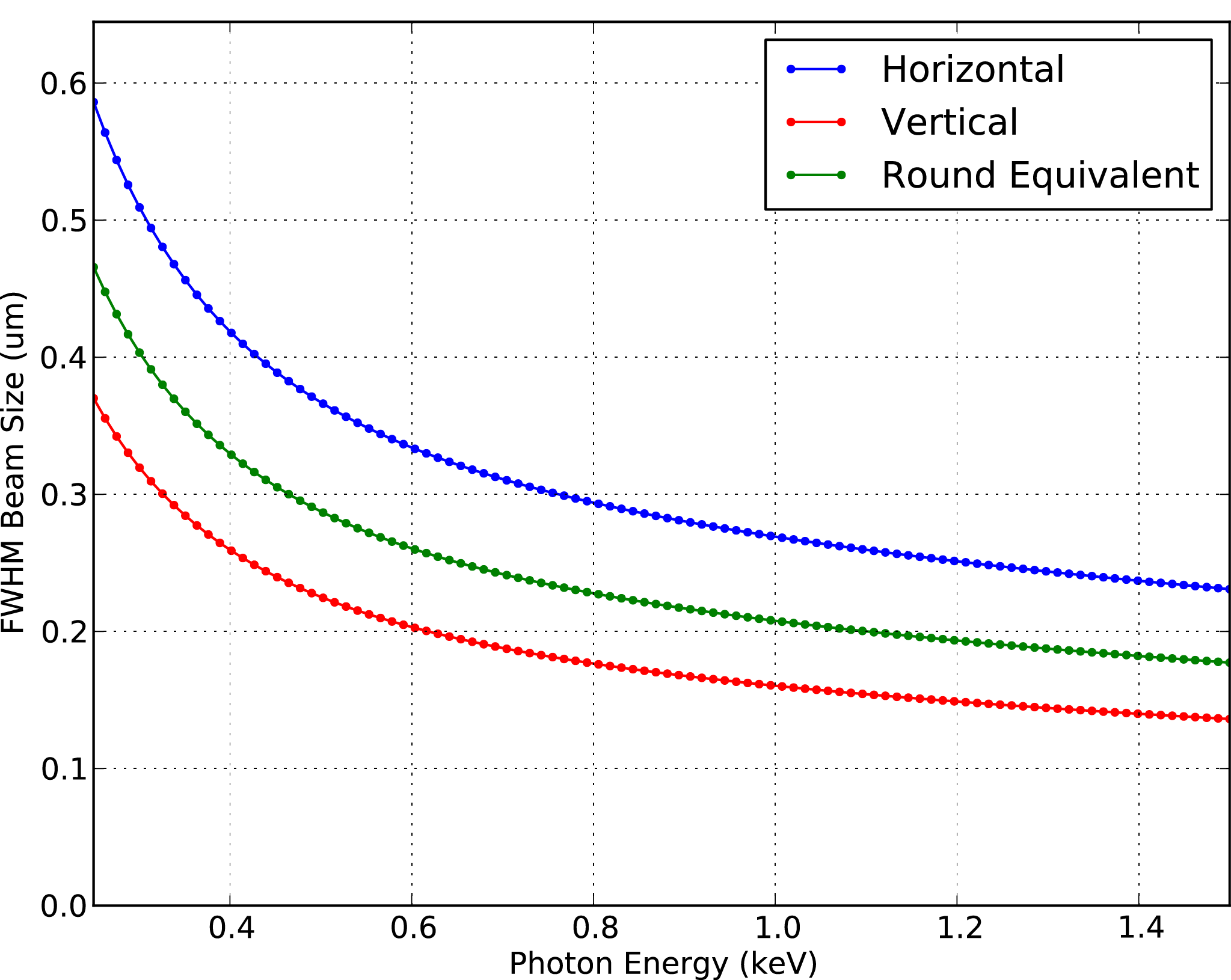}
    \caption{Flower two.}
  \end{minipage}
  \caption{ Exemplary smallest achievable focus spot sizes at both IPs in TMO, achieved from calculations based on start-to-end simulations for LCLS II with 20 pC bunch charge.}
\label{fig:focus1}
\end{figure}

The essential X-ray focusing optics in TMO are KB-mirrors. Each of the two KB-systems are tailored to the applications of the corresponding focus spot, also called interaction points. Commonalities are the two different reflection zones, one of a B\textsubscript{4}C coating and bare Si to facilitate a maximum transmission over the energy range covered by TMO. The energy range of TMO is mostly defined by the material dependent reflection cut off with the given incident angle. The two different coatings of the mirrors can be reached with a translation of the mirror Si-substrate. Such a translation also allows for different deflection areas in case of mirror contamination. The incident angle with respect to the mirror surface are 14 mrad and 21 mrad for the first and second pair of KB mirrors, respectively. These incident angles allow for an X-ray energy range of 250 - 2200 eV on the first focus spot and an X-ray energy range of 250 - 1400 eV for the second focus spot.  The first pair of mirrors are bendable in a plain-elliptical geometry which enables the foci to be varied dynamically along the first focus spot as well as to compensate the aberration of the second pair of mirrors which is not bendable. Fig. \ref{fig:KBO} shows the TMO focus scheme and geometry and Fig. \ref{fig:focus1} depicts the focus parameters for the different photon beam energies. 
\subsection{Interaction point one (IP1)}

At the first focus spot of the TMO instrument is the Next generation Atomic, Molecular, and Accelerator Science and Technology Experiments (NAMASTE) and it is optimized for performing high energy, high-resolution, time-but also angular-resolved charged particle measurements. It accepts highly standardized modular endstations. The NAMASTE environment offers the possibility to install modular stations (roll in and out) which can be set up, aligned and commissioned outside the hutch and installed at the first TMO focus spot.   
\subsubsection{LAMP}
The LAMP instrument is a soft X-ray endstation for high-field physics and ultrafast science experiments. LAMP contains a high-resolution double sided electron-ion coincidence Velocity Map Imaging (VMI) spectrometer specifically designed for use in the LAMP endstation. This detects ions and/or electrons allowing for simultaneous and, under certain conditions coincident measurements of all charged particles generated by photoionization of a single atom/molecule.\cite{Li2021} More information on LAMP can be found here~\cite{Osipov2018}.  
\subsubsection{Magnetic Bottle Spectrometer}
The TMO magnetic bottle electron spectrometer (MBES) features a 2.1 meter flight tube and an electrostatic lens to retard high energy photo-electrons, for improved kinetic energy resolution. The spectrometer has been integrated into the LAMP experimental chamber \cite{Osipov2018} but is designed to enable flexible integration with different endstations. An inhomogeneous magnetic field of strength ~0.5 T is created at the interaction region by a neodymium permanent magnet mounted with a soft iron cone which acts as a magnetic shunt. A flat copper plate is attached to the nose cone, which enables the application of a uniform electric field across the interaction region by applying a voltage to the magnet. The tip of the iron cone sits 3 mm from the interaction region, opposite a copper nose also situated 3 mm cone from the interaction region. The aluminum flight tube has a 76 mm outer diameter and is mounted in vacuum. It is wrapped along its length by Kapton-insulated 22 AWG copper wire, through which a current of ~1 A is flowed to create a uniform solenoid magnetic field of ~1 mT. The electrons are detected by a microchannel plate detector coupled to an anode at the end of the flight tube. The detector is mounted opposite the flight tube in a four-way cross, with the remaining two ports occupied by an electrical feedthrough flange and a turbomolecular pump. 
The electrostatic lens consists of two pairs of plates and sits 19 cm from the interaction region. At this point, the magnetic field lines created by the combined fields of the permanent magnet and solenoid are almost fully parallel. To retard high kinetic energy electrons, the set of plates closest to the interaction region is held at ground while the other set of plates holds the retardation voltage. The flight tube is also held at the retardation voltage to ensure the electrons experience field-free drift after interacting with the electrostatic lens.
\subsubsection{coaxial-VMI}
The coaxial VMI endstation provides an electron VMI spectrometer where the ionizing laser source propagates along the symmetry axis of the spectrometer. This coaxial VMI provides a unique 2-dimensional projection of the 3-dimensional electron momentum distribution. The coaxial imaging technique has been demonstrated in experiments at the Linac Coherent Light Source with both soft X-ray and infrared laser pulses.  More information can be found here \cite{Li2018}.
\subsubsection{Angular resolved high-resolution electron spectrometer}\label{MRCOFFEE}
\begin{figure}
\caption{Axial view of the multi-resolution electron spectrometer array detector along the direction of X-ray propagation, showing the 16 azimuthal eToFs MRCOFFEE image taken from\cite{MRC}}
\includegraphics[width=1\textwidth]{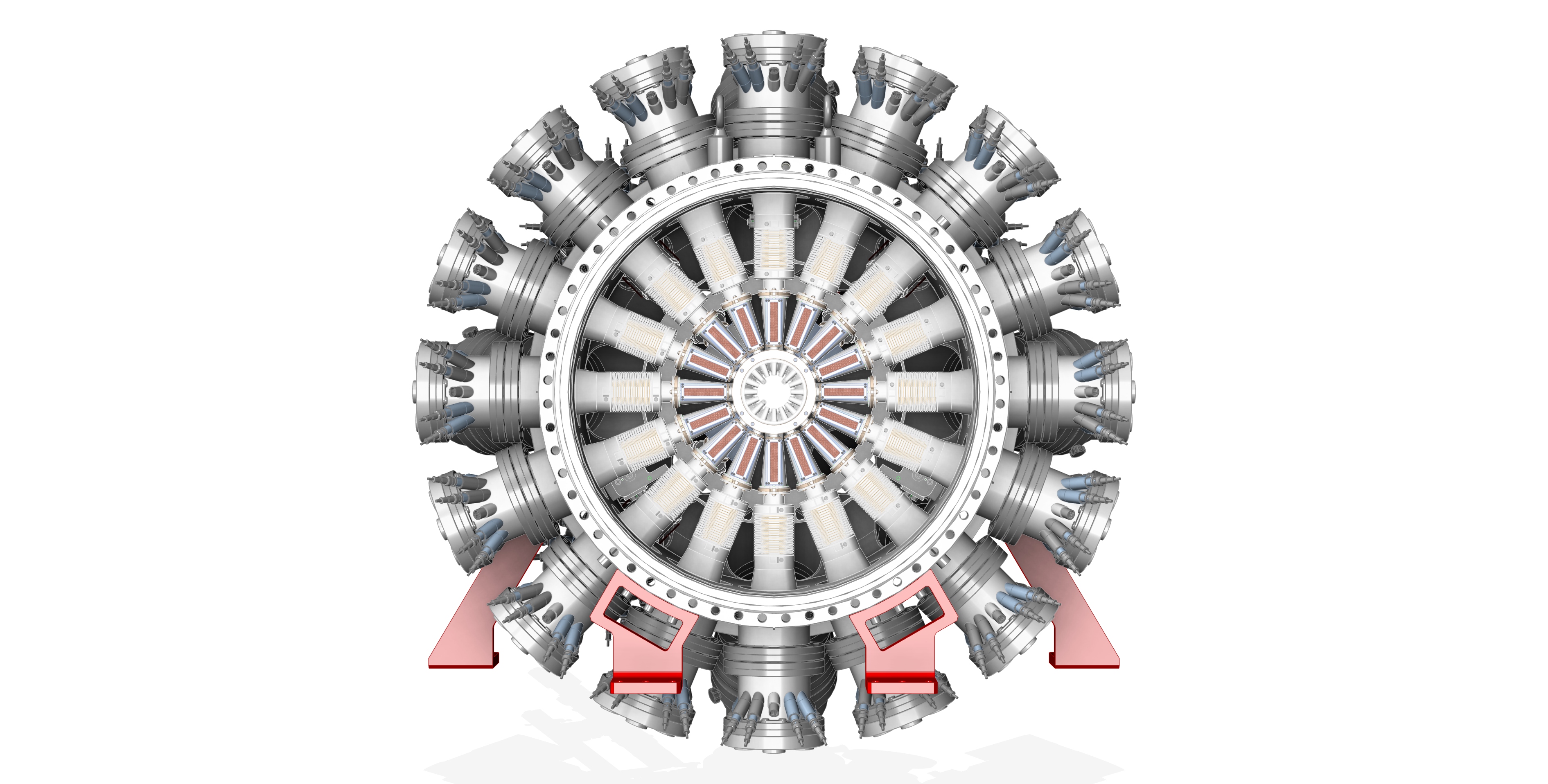}
\label{fig:MRC}
\end{figure}

Based on the design of the P04-Beamline \cite{Viefhaus2013a} at PETRA III diagnostic unit (so called 'cookiebox'), \cite{Durr2016, Hartmann2018} TMO will have an in-house developed experiment and diagnostic endstation, called Multi-resolution Cookiebox Optimized for the Future of Free Electron laser Experiment - MRCOFFEE \cite{MRC}. This Multi-resolution spectrometer endstation is an angle-resolving array of 16 electron time-of-flight spectrometers that allow wide and adjustable energy acceptance windows (see Fig. \ref{fig:MRC}). By interleaving detector retardations, it enables simultaneous angle-resolved photo-electron and Auger electron spectroscopy. 
The spectrometer array is available for: multiple-edge high-resolution photoelectron and Auger electron spectroscopy,   spectral-polarimetry measurements as well as polarization sensitive attosecond resolving temporal characterization of general LCLS-II pulses. This multi-polarization and multi-color spectral experimental endstation and/or diagnostic unit also has an energy resolution better than the expected seeding spectrum and the SASE spectral features.
\subsubsection{Future upgrades for IP1}  
Provisions are in place to accommodate additional experimental setups that would further increase the experimental capabilities at the TMO instrument. Possible future upgrades can be, but not limited to soft X-ray high repetition rate imaging and/or a photon spectrometer. 
\subsection{Interaction point two (IP2)}
\begin{figure}
\centering
\includegraphics[width=0.6\textwidth]{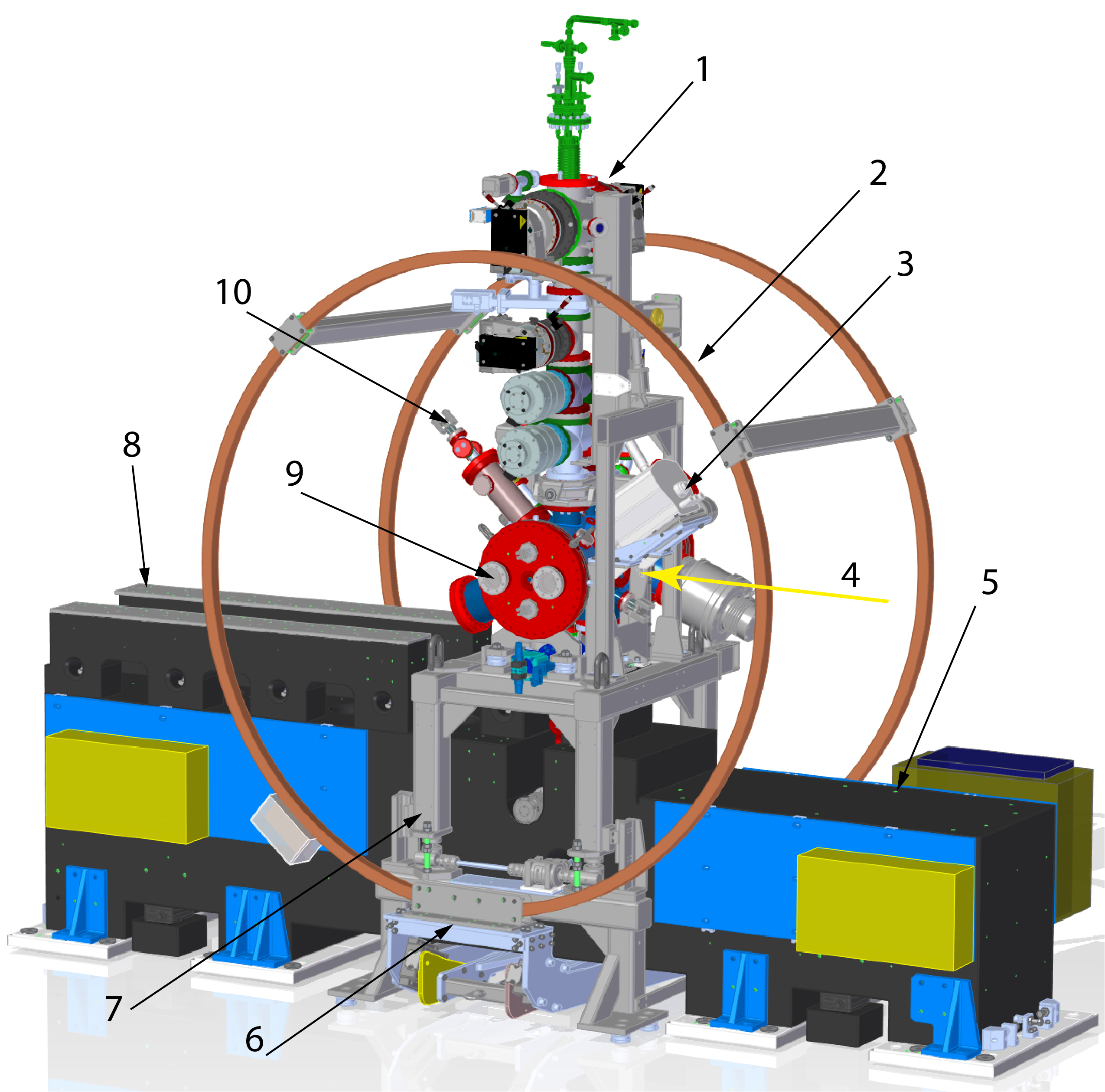}
\label{fig:DREAM}
\caption{Overview of the DREAM endstation (IP2). Shown are the 4 skimmer stage jet (1), the Helmholtz coils (2), long-range microscope (3), the X-ray trajectory (4), the KBO position (5), the Helmholtz coil mover (6), main chamber vertical motion stand (7), DREAM diagnostics position (8), optical laser in$/$out-coupling (9), and sample paddle (10) }
\end{figure}

A new Dynamic REAction Microscope (DREAM) is positioned at the second interaction point. DREAM houses a well-defined geometry and Cold Target Recoil Ion Momentum Spectroscopy (COLTRIMS) type spectrometer \cite{Dorner2000, Ullrich2003} as a standard configuration to accommodate extreme vacuum, sub-micron focus spot size, and target purity requirements dictated by the pump-probe class of coincidence experiments, while accumulating data on an event-by-event basis at repetition rates in excess of 100 kHz, utilizing the LCLS-II capabilities (see Fig. \ref{fig:DREAM}. The Photon-fluence in DREAM will reach over 10\textsuperscript{21} photons/cm\textsuperscript{2} in combination with the superconducting Linac, while with the copper accelerator it will be over 10\textsuperscript{22} photons/cm\textsuperscript{2} at 120 Hz. 
The new DREAM instrument enables sophisticated coincidence measurement schemes for kinematically complete experiments at each time step of an evolving reaction. This experimental approach, known as a “molecular reaction microscope” will enable the complete spatial reconstruction of the excited-state charge transfer and subsequent dissociation at each time step for a fixed-in-space molecular orientation. This is a powerful new approach for visualizing a broad range of excited-state molecular dynamics. With the high repetition rate of LCLS II, with up to 1 MHz, it will push the boundaries of coincidence measurements. \cite{Jahnke2020, Li2021}\\
The standard detectors for this COLTRIMS endstation are two 120 mm state-of-the-art Hexanode delay-line detectors and a setup that allows tailored spectrometers sizes (length and diameter). A more detailed description of the DREAM endstation will be published elsewhere. 
\subsubsection{Future upgrades for IP2}  

To overcome the limitations of a delay-line detector, SLAC is developing a spatial and time resolving front-end ASIC detector. This time resolving pixelated detector is called TIXEL and would provide the time of arrival and time over threshold. This new develoment is supposed to work without a MCP in front of the TIXEL detector. With such a pixelated charged particle detector based on silicon technology and with $>$ 300 hits at a repetition rate of up to 1 MHz it will be possible to perform for the first time massive coincidence experiments. Since the number of combinations grows exponentially with the number of hits this detector it will be possible to detect and analyze large densities of hits with high repetition rate. Possible further future upgrades can be, but not limited to, compact jet system and/or aerosol environment.
\subsection{FEL capabilities}\label{FEL}

The soft X-ray line of LCLS-II is equipped with 21 variable gap undulators. The soft X-ray (SXR) undulator can be operated with either the superconducting LCLS-II linac, or the normal conducting linac. In the first case the design energy range is between 250 eV and 2.5 keV, while in the second case the undulator can generate FEL radiation up to several keV.

The SXR line will leverage several advanced FEL capabilities developed with LCLS:

- self-seeding will enable the generation of narrow bandwidth radiation from the undulator \cite{ratner2015experimental}

- XLEAP-II will generate attosecond pulses with tens of uJ of pulse energy and duration down to 200 as using enhanced SASE \cite{duris2020tunable,zholents2005method}

- two-color operation with the split undulator method will allow the generation of two pulses with wide energy separation (determined by the undulator tuning range) and delay control up to 1 ps \cite{PhysRevLett.110.134801}

Combining the XLEAP-II modulators with the two-color mode will enable X-ray pump/X-ray probe experiments with sub-fs resolution.
All three capabilities can be realized at high repetition rate using the superconducting LCLS-II linac.

Ongoing R\&D aims at enabling femtosecond shaping of the electron beam to smoothly control the pulse duration \cite{marinelli2016optical}, advanced self-seeding to improve pulse-to-pulse stability \cite{PhysRevLett.110.134801} and double chirp/taper operation \cite{zhang2019double} for high power two-color attosecond operation and single TW-scale pulses. 
\subsection{Optical laser capabilities}

Optical excitations of samples in the target chambers are provided by synchronised, femtosecond laser pulses delivered from lasers that are situated in the NEH laser hall. Currently, these pulses are generated in Ti:Sapphire Chirped Pulse Amplification (CPA) systems, providing up to 20 mJ, $~$40 fs laser pulses at 800 nm and 120 Hz. Ultimately an in-house developed Optical Parametric Chirped Pulse Amplification (OPCPA) system will provide 800 nm, 100 kHz, 35 W, $<$20 fs pulses. Each endstation has an associated laser delivery setup, consisting of a series of modules for conditioning and manipulating the drive laser, with the possibility of generating pulses ranging from UV wavelengths to THz frequencies. An emphasis has been placed on the incorporation of numerous control and diagnostic points throughout the laser delivery system, to monitor the system performance, improve reliability, and allow for greater remote control of laser properties. Laser in-coupling is tailored to the specific needs at each interaction point. For example, in-coupling to the DREAM endstation is designed to produce a $<$10 $\mu$m FWHM focal spot, whereas two in-coupling paths are incorporated into IP1, one for UV-visible wavelengths and a separate path for Near InfraRed (NIR)-Mid InfraRed (MIR) wavelengths. In order to determine a shot-by-shot time-of-arrival of the laser pulse relative to the X-rays, a small portion of the laser pulse is used to perform a cross-correlation with the X-rays.\cite{Muhammad2021} This enables re-sorting of data into appropriate time bins, improving the temporal resolution of measurements. The incorporation of a distributed, laser-based timing system into these cross-correlation measurements will provide a reference clock that is accurate to the $\sim$1 fs level, and is expected to enable temporal resolution below 10 fs.
\subsection{Sample delivery}

The new TMO instrument provides several different gas phase and aerosol delivery options as well as the option to host condensed and solid phase targets. A new remote controllable and automated sample delivery system is part of TMO. Each interaction point has its own gas delivery environment, roughing line system and exhaust lines to make sure both interaction points can be operated completely independent. TMO can provide an Even-Lavie valve \cite{Even2015}, high pressure Parker valve, a tailored CW jet, aerosols and effusive sample gas delivery systems, flow cell, as well as sample stages for solid targets. As a future upgrade, a high repetition rate pulsed gas jet is foreseen ($<$ 5kHz). To keep the high transmission of the mirror system and avoid contamination, liquid jets are not foreseen at TMO. 
\subsection{Diagnostics}

The TMO instrument hosts a suite of different destructive and non-destructive diagnostics to characterize the X-rays. Beside standard beam positioning monitors along the beam trajectory (see Fig. \ref{fig:TMO}) each focus spot in TMO has its own WaveFront Sensor (WFS) using the fractional Talbot effect. \cite{Liu2020} Given the high repetition rate of the FEL each WFS will be operated in an average mode. Each WFS, however, can be used with the full beam transmission. The destructive nature of this kind of measurement makes the simultaneous use of both focus spots with both WFS inaccessible. To measure the beam transmission downstream of every mirror system, TMO has power meters after each X-ray optic.\cite{Heimann2019PM} To measure the position of the X-ray beam on each mirror and for non-invasive, pulse by pulse normalization a Fluorescence Intensity Monitor (FIM) has been developed \cite{Heimann2019FIM} and will be available as a standard TMO diagnostic on each TMO mirror. A photon spectrometer as a pulse diagnostic is under development (see section \ref{MRCOFFEE}). To measure the average, absolute and pulse-resolved photon flux of the FEL beam, two Gas Monitor Detectors (GMD) are implemented before and after the gas attenuator. \cite{Tiedtke2008, Tiedtke2014} \\
\begin{table}[]
\begin{center}
\centering
\caption{Parameters and capabilities of the TMO instrument.}
\label{table:2}
\begin{tabular}{lcc}
\hline
TMO parameter                   & IP 1                   & IP 2                   \\ \hline
Energy range (eV)               & 250 - 2200             & 250 - 1400             \\
Mirror incidence angle (mrad)   & 14                     & 21                     \\
Mirror coating                  & \multicolumn{2}{c}{Si or B\textsubscript{4}C}     \\
Smallest focus ($\mu$m)             & \textless 1.5          & \textless 0.3          \\
Flux, up to (photon/cm\textsuperscript{2})      & 10\textsuperscript{21} & 10\textsuperscript{22} \\
Mirror transmission ($\%$)      & ~ 80                   & ~ 60                   \\
Pulse duration (fs)             & \multicolumn{2}{c}{0.3 - 100}                   \\
Repetition rate X-rays (kHz)           & \multicolumn{2}{c}{up to 929}                         \\
Target sample                   & \multicolumn{2}{c}{gas jet, aerosol, solid}      \\
Gas sample delivery             & \multicolumn{2}{c}{CW and puled valves}         \\
Gas sample delivery temperature (K) & \multicolumn{2}{c}{10 - 650}                \\
Optical laser source            & \multicolumn{2}{c}{OPCPA}                       \\
Peak intensity on  Sample (W/cm\textsuperscript{2})\footnote{ at 100kHz and 800 nm}     & 10\textsuperscript{14}                & 10\textsuperscript{14}   \\
Repetition rate OPCPA (kHz)\footnote{later upgrade to MHz}     & \multicolumn{2}{c}{100} \\
Optical laser wavelength ($\mu$m)   & 0.2 - 10               & 0.2 - 5              \\
Optical laser focus spot size($\mu$m)\footnote{ at 800 nm}    & 25               & $<$ 10            \\
\hline
\end{tabular}
\end{center}
\end{table}
\section{TMO science}

With the focus of TMO on ultrafast X-ray atomic and molecular physics, we have a powerful tool to tackle new challenges: Charge migration, redistribution and localization as well as symmetry break down and chirality, even in simple molecules, are not well understood at the quantum level. These fundamental phenomena are central to complex processes such as photosynthesis, catalysis, and bond formation/dissolution that govern all chemical and biochemical reactions. Ultrafast soft X-rays at high repetition rate from LCLS-II will provide qualitatively new probes of excited-state energy and charge flow and how they work in simple and complex molecular systems. The following section will give an exemplary overview of science cases enhanced by TMO and LCLS II.
\subsection{Investigation of ultrafast processes with attosecond soft X-ray laser pulses}
As attosecond science plays a central part to our LCLS-II scientific mission, TMO gives the possibility to follow fundamental phenomena from attosecond charge migration involving electronic dynamics and correlation to femtosecond charge transfer involving nuclear rearrangement. Understanding charge dynamics at the microscopic level is essential not only to achieving fundamental insights into these phenomena, but also, on a longer-term perspective, for improving applications relying on charge motion and separation, from artificial photosynthesis and molecular electronics to photocatalytic and photovoltaic devices. The continuous tunability and orders-of-magnitude pulse energy increase (compared to any previous attosecond source) produced by enhanced SASE operation at the LCLS enable a suite of nonlinear spectroscopies which are, at the moment, unavailable elsewhere. Therefore, the new capabilities of TMO and LCLS II can be used to perform measurements demonstrating the control and observation of coherent electronic motion on its natural attosecond timescale, and explore how it affects the subsequent motion of the nuclei to drive chemical change.
\subsection{Ultrafast dynamics in chiral molecules}
With the new features of LCLS II, i.e. a polarization-tunable undulator, sub- femtosecond pulses, multibunch modes, up to 1 MHz repetition rate (see Section \ref{FEL}), as well as the new TMO instrument with the above described capabilities, SLAC has opened the door for unprecedentedly deep insights into chirality.  It will finally be possible to observe how chiral systems form, restructure and how their functionality can be understood or even controlled. Investigating the origin of chirality with TMO will build bridges between the macro-biological level, chemical dynamics and the fundamental spin properties of matter. \cite{Pitzer2017, Hartmann2016, Ilchen2021}   
\subsection{Ultrafast photochemistry}
In combination with a high repetition rate soft X-ray source, the TMO hutch is well-equipped for the investigation of ultrafast photochemistry. The high average X-ray flux will allow the measurement of orders of magnitude weaker signals using already established time-resolved experimental methods. \cite{wolf2017, mcfarland2014} Furthermore, the high repetition rate will enable novel experimental techniques based on coincidence detection, allowing for example to associate spectral signatures to specific fragmentation patterns. Such techniques will help in differentiating multiple parallel photochemical reaction channels and in detecting reactive processes with small quantum yields, which can nevertheless be of significant interest, e.g. for organic chemistry.
\subsection{Future flagship capabilities}
With the new TMO beam line and the two X-ray focus spots in line, TMO will be able to characterize the substructure of SASE pulses and correlate it with the time-resolved study of light-matter interactions for charged particle spectroscopy. It has been shown that the temporal sequence of spikes can crucially influence the underlying ionization pathways.\cite{Li2021} Moreover, at very short pulse durations, SASE pulses are spectrally broad and therefore, each pulse possesses a specific temporal and spectral profile, which ultimately determines the non-linear excitation process in all FEL-related photon energy regimes. For the SASE pulse characterization, we plan to implement the method of ''angular  streaking'' at TMO IP1 with MRCOFFEE.\cite{Hartmann2018, MRC} Enabled by this novel technique, we will spectrally and temporally characterize the SASE pulses at the IP1, incl. their single-shot duration with sub-fs resolution, with the full time-energy information for each incoming XFEL shot. \cite{Driver2020}
The DREAM endstation in IP2 will be a superior endstation to perform linear-, high field-, coincidence-, covariance-, and Coulomb explosion imaging-experiments with full analysis of the X-ray pulse spectrum. It will lead to a complete analysis of the underlying electron dynamics from the birth of a (photo-) electron to the early state of charge migration, charge transfer, to nuclear motion and chemical reactions. 
\section{TMO first results}
\begin{figure}
\centering
\includegraphics[width=0.4\textwidth]{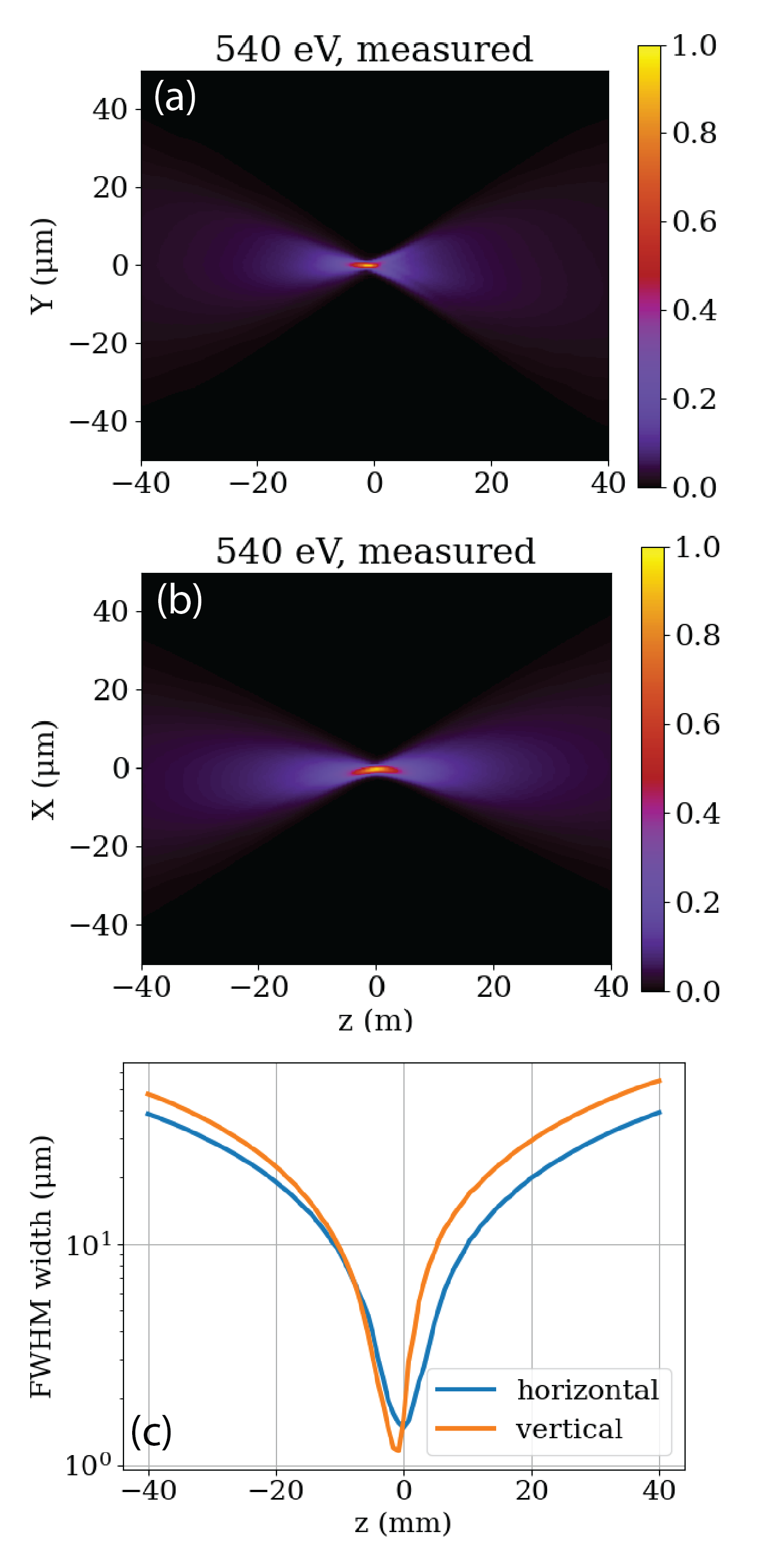}
\label{fig:WFS}
\caption{First wave front measurements at the IP1 focus spot of TMO. Shown are the horizontal measured line-outs along the beam (a), the vertical measured line-outs along the beam (b), as well as the FWHM for both (c). }
\end{figure}

\begin{figure}
\centering
\includegraphics[width=0.95\textwidth]{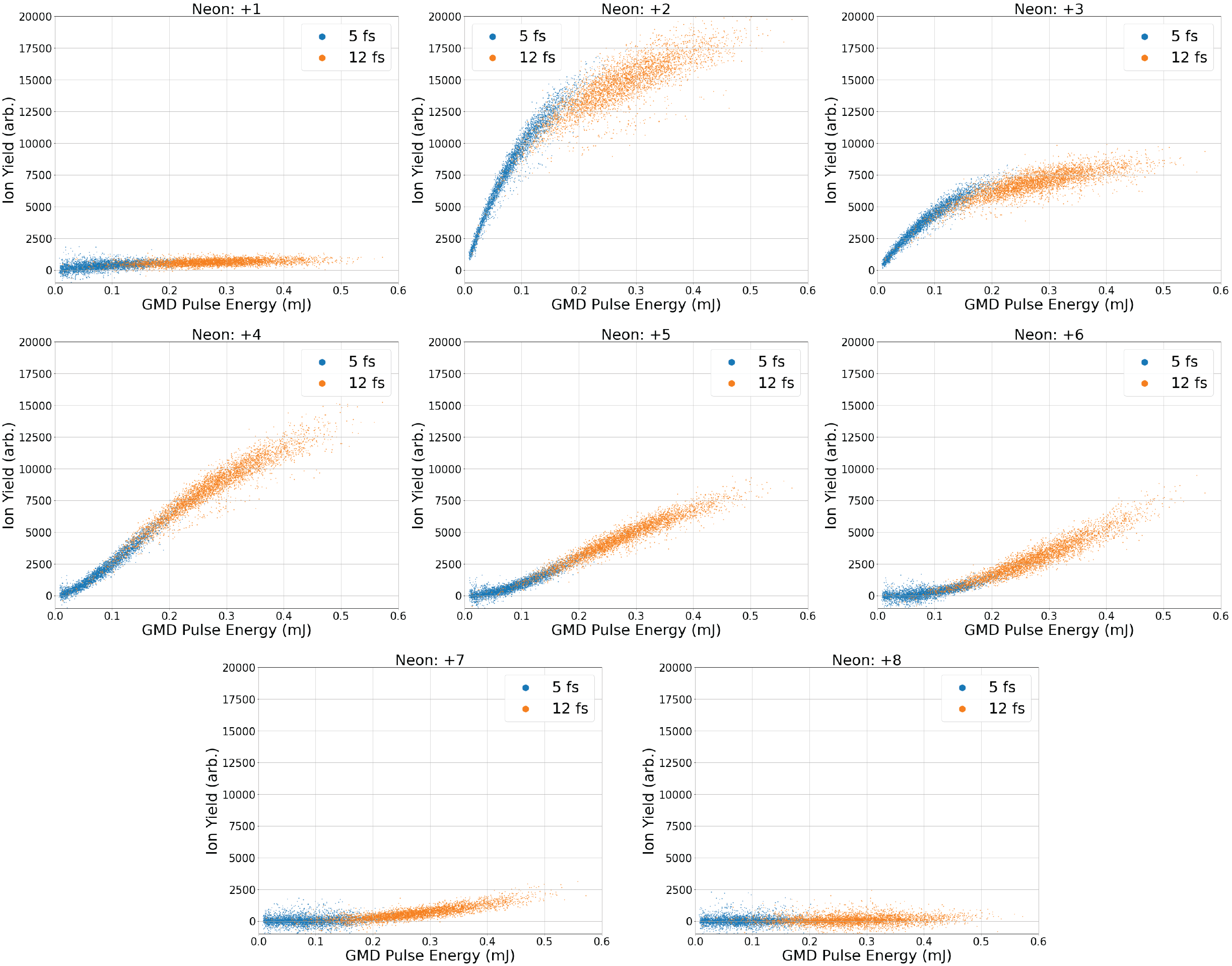}
\caption{Ratio of charge state yield versus XGMD pulse energy for 5 fs (blue) and 12 fs (orange) electron bunch duration and at a photon energy of 1030 eV.Pulse energies are measured in the XGMD upstream of the target.The charge states 9+ and 10+ were not observed in the single shot spectra. The ratios are in good agreement with the literature for this photon energy.\cite{Young2010}}
\label{fig:NEcharge}
\end{figure}

\begin{figure}
\centering
\includegraphics[width=0.5\textwidth]{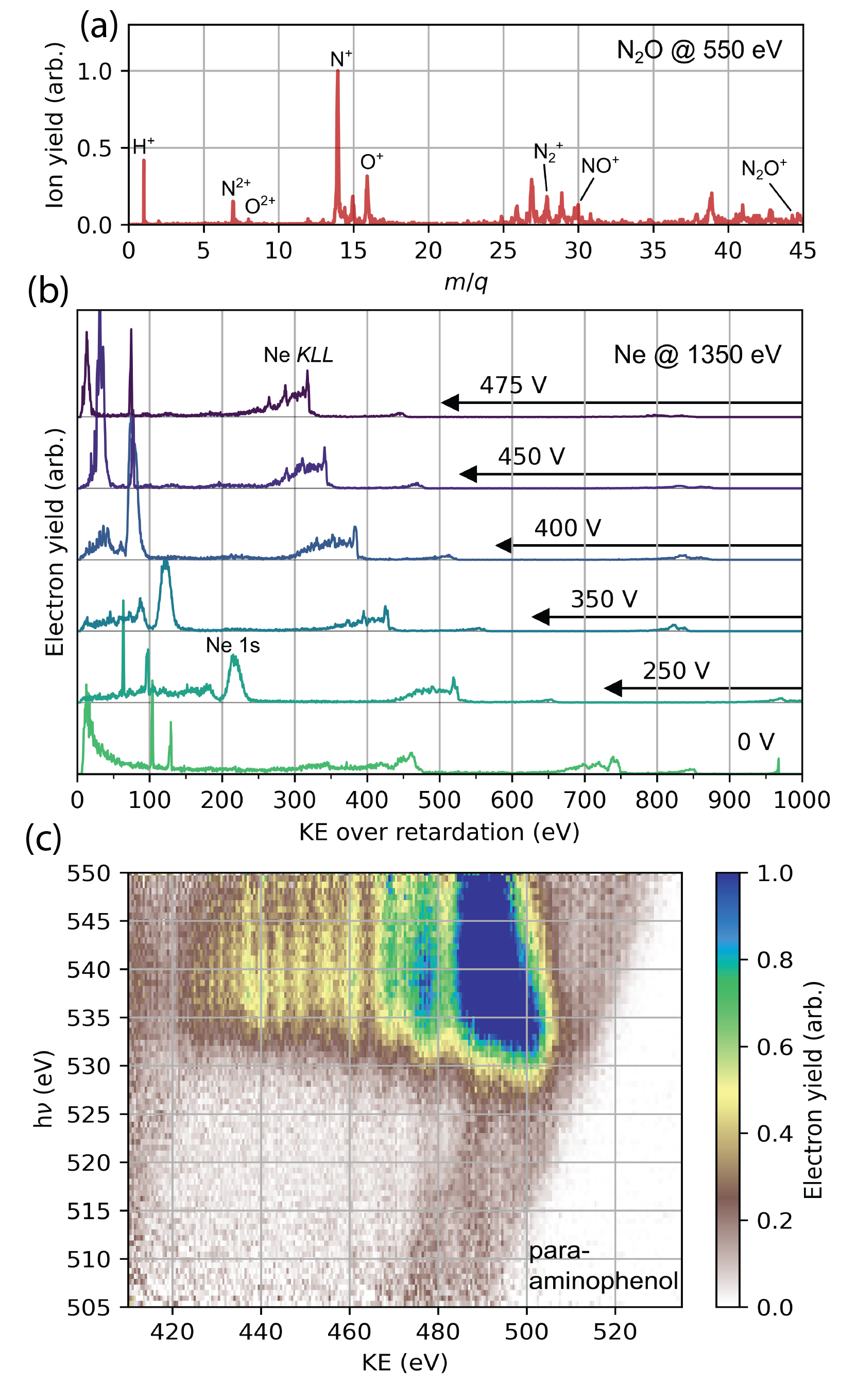}
\label{fig:MBES}
\caption{ Ion mass-to-charge spectrum of N\textsubscript{2}O irradiated by 550 eV x-ray pulses recorded with the magnetic bottle spectrometer (a). Photoelectrons produced by ionization of Ne atoms by 1.35 keV x-rays, recorded at different retardation voltages applied to the electrostatic lens. As the retardation voltage is increased, the manifold of three primary peaks which form the Ne KLL Auger-Meitner emission spectrum \cite{KRAUSE1970} becomes more sharply defined (b).
Resonant Auger-Meitner electron emission map of gas-phase 4-aminophenol, recorded with the MBES. Incoming photon energy is plotted on the y-axis and electron kinetic energy is plotted on the x-axis. At photon energies corresponding to the manifold of O 1s $->$ valence excitations in the molecule (~535 eV) there is a resonant enhancement in the Auger-Meitner electron yield. The emission converges to normal KLL Auger-Meitner decay above the O K-edge at ~540 eV. The highest kinetic energy photoelectrons are produced by x-ray ionization of the valence shell and show the expected linear dispersion with photon energy (marked by dotted line)(c).}
\end{figure}

The TMO hutch opened the X-ray stoppers for the first photons October 09th, of 2020 and had its first successful experiment only one month later. At first, TMO conducted a series of high-field experiments to demonstrate its main new capabilities. Most of the results of this first series of experiments will be published elsewhere. Here we present some selected results. 
The first wave front measurement was made during the first week of commissioning. Fig. \ref{fig:WFS} shows the line outs of the horizontal and vertical focus, as well as the FWHM for both focus orientations. The residual astigmatism and some minor coma are expected to be removed during future mirror bender optimizations. 
The first set of experiments performed at TMO has been a series of high-field physics to demonstrate but also make sure the tight focus spot in IP1 can be reached and the high fields for non-linear science can be achieved. The series of performed experiments were, but not limited to, a set of high field and double core hole experiments on Ne and N\textsubscript{2}O, two photon non sequential double ionization experiment, but also two photon absorption experiments using the new $\omega$/$2\omega$ capabilities of the new variable gap undulators and the self-seeding/delay stage to use the split undulator configuration to achieve phase stable pulse pairs. \\
Here we present data showing Ne ionization states reached with two different photon energies and photon pulse durations using the LCLS slotted foil\footnote{Note: An emittance spoiling foil is only available on the copper accelerator site and is not available for high repetition rate operating superconducting accelerator}.\cite{Emma2004, Ding2015} Figure \ref{fig:NEcharge} shows the experimental ion charge state yield versus XGMD measured pulse energy for two different pulse durations, 5fs and 12 fs (FWHM) pulses, at a photon energy of 1030 eV. At this photon energy we expect different ionization mechanisms - valence ionization, inner-shell ionization and ionization in the regime far above all edges of all charge states of Neon. Once we have reached the ground-state configuration of berillium-like neon (Ne\textsuperscript{6+}), the binding energy of the core-level electrons is greater than the incident photon energy, so only valence ionization is possible. However, at these pulse durations we did not observe Neon\textsuperscript{9+} nor Neon\textsuperscript{10+}. With a spot size of less than 1 $\mu$m we should be able reach the needed fluence to gain the state of a hydrogen-like neon to get the not observed high charge states of Neon\textsuperscript{9+} and Neon\textsuperscript{10+}. While short pulse durations like these will be great to observe shake and Auger-Meitner processes here the durations could be to short for multi-photon processes to reach the high charge states.\cite{Young2010}  Also at this moment of time a sub-optimal focus spot could not be excluded. A set of modeling and simulations will help to further analyse this data.
Further results are from the MBES. Figure \ref{fig:MBES} illustrates the performance of the TMO MBES. The top panel shows the ion mass-to-charge spectrum recorded following x-ray ionization of nitrous oxide at a photon energy of 550 eV. To collect ions, a voltage of +5 kV is applied to the copper plate mounted to the permanent magnet and the lens retardation voltage is held at 0 V. The middle panel shows the electron kinetic energy spectrum of photoemission from neon gas at an x-ray photon energy of 1.35 keV, with different retardation voltages applied to the electrostatic lens. The retardation voltage is shown on the right hand side of the figure. Increasing the retardation voltage slows the photoelectrons and improves kinetic energy resolution. The manifold of peaks at ~800 eV kinetic energy corresponds to Ne KLL Auger-Meitner emission, which is dominated by three primary features each separated by ~30 eV.\cite{KRAUSE1970} At higher retardation voltages, these three peaks become clearly resolved.
\section{Conclusion}

LCLS and LCLS II produce high-flux few to sub-femtosecond X-ray pulses, yielding unprecedented X-ray intensities. The TMO instrument takes advantage of the pulse properties to perform high-power soft X-ray experiments in a wide spectrum of scientific domains. The instrument provides users with a variety of endstations, spectrometers and other components for the utmost flexibility in experimental layouts and signal detection schemes. The latest updates and more details about the TMO instrument can be found on the following website: https://lcls.slac.stanford.edu/instruments/neh-1-1
\section{Facility access}

LCLS instruments are open to academia, industry, government agencies and research institutes worldwide for scientific investigations. There are two calls for proposals per year and an external peer-review committee evaluates proposals based on scientific merit and instrument suitability. Access is without charge for users who intend to publish their results. Prospective users are encouraged to contact instrument staff members to learn more about the science and capabilities of the facility,and opportunities for collaboration.
\section{Acknowledgements}

This research was carried out at the Linac Coherent Light Source (LCLS) at the SLAC National Accelerator Laboratory. SLAC National Accelerator Laboratory, is supported by the U.S. Department of Energy, Office of Science, Office of Basic Energy Sciences under Contract No. DE-AC02-76SF00515. We thank Gregory Stewart and Terry Anderson for helping with Fig. \ref{fig:LCLS} $\&$ \ref{fig:KBO}. The authors want to thank the LCLS TMO instrument advisory panel for their constant support and fruitful discussions. We furthermore acknowledge the entire LCLS, LCLS II, L2SI staff, and also the SLAC vacuum, metrology and machine shops staff for their assistance during the design, construction, and commissioning of the TMO instrument. 
\bibliographystyle{iucr}
\bibliography{TMO.bib}
\end{document}